\documentclass[12pt]{article}
\usepackage{amsmath,amssymb,amsfonts,commath}
\usepackage{graphicx}
\usepackage{hyperref}
\hypersetup{colorlinks,linkcolor={blue},citecolor={blue},urlcolor={red}} 
\usepackage{cite}
\usepackage[english]{babel}
\usepackage{bm}
\usepackage[utf8]{inputenc}
\usepackage{listings}
\usepackage[T1]{fontenc}
\usepackage{lmodern}
\usepackage{caption}
\usepackage{floatrow}
\usepackage{subcaption}
\usepackage{mathrsfs}
\usepackage{enumerate}
\usepackage[normalem]{ulem}
\usepackage{color}
\usepackage{array}
\newcommand{\deltafun}{\delta\!}
\newcommand{\erf}{\rm erf\!}

\newcommand{\imgscale}{0.35}

\def\beq{\begin{equation}}
\def\eeq{\end{equation}}
\def\be{\begin{equation}}
\def\ee{\end{equation}}
\def\bea{\begin{eqnarray}}
\def\eea{\end{eqnarray}}

\begin{document}

\title{\hfill\mbox{\small}\\[-1mm]
\hfill~\\[0mm]
       \textbf{Predictions for the Leptonic Dirac CP-Violating Phase}        }
\date{}
\author{\\[1mm]Lisa L.~Everett$^{1\,}$\footnote{E-mail: {\tt
leverett@wisc.edu}}~,~Raymundo~Ramos$^{2\,}$\footnote{E-mail: {\tt raramos@gate.sinica.edu.tw}}~,~Ariel B. Rock$^{1\,}$\footnote{E-mail: {\tt arock3@wisc.edu}}~,~\\and Alexander J.~Stuart$^{3,4\,}$\footnote{E-mail: {\tt astuart@ucol.mx}}\\[1mm]
 \textit{\small $^1$Department of Physics, University of Wisconsin,}\\
  \textit{\small Madison, WI 53706, USA}\\[3mm]
  \textit{\small $^2$Institute of Physics, Academia Sinica,}\\
  \textit{\small Nangang, Taipei 11529, Taiwan}\\[3mm]
 \textit{\small $^3$Facultad de Ciencias-CUICBAS, Universidad de Colima,}\\
  \textit{\small C.P.~28045, Colima, Mexico}\\[3mm]
  \textit{\small $^4$ Dual CP Institute of High Energy Physics,}\\
  \textit{\small C.P.~28045, Colima, Mexico}\\[3mm]
  }

\maketitle

\vspace{-0.5cm}

\begin{abstract}
\noindent 
We explore the theoretical constraints on the observable parameters of neutrino mixing on predictions for the leptonic Dirac CP-violating phase within a well-studied class of simple theoretical models that includes a single source of CP violation due to charged lepton corrections. The approach guarantees that a physically meaningful prediction for the most likely values for the leptonic Dirac CP-violating phase is obtained.
\end{abstract}

\thispagestyle{empty}
\setcounter{page}{0}
\vfill
\newpage

\newpage


\section{Introduction}
\label{sec:intro}

The confirmation of a non-zero and sizable reactor mixing angle~\cite{dayabay,reno,doublechooz} has opened the window to detecting CP violation in the lepton sector through the direct measurement of the Dirac CP-violating phase $\delta$ of the Maki-Nakagawa-Sakata-Pontecorvo (MNSP) lepton mixing matrix, $U_\mathsf{MNSP}$~\cite{Pontecorvo:1957cp,Maki:1962mu}; see also the PDG review~\cite{pdg}.  There are already experimental  hints that its value may exist around $\delta \sim \pm \pi/2$ from the T2K~\cite{c5} and NO$\nu$A~\cite{c6} collaborations.  Additionally, input from current global fits favor  $\delta \sim-\pi/2$  at 3$\sigma$~\cite{Esteban:2018azc,nufitweb,c1,Capozzi:2018ubv}.  In either case, the impending confirmation of the value of this CP-violating phase forces physicists to revisit theories of its possible origin which can explain its measured value. 
The arguably most popular approach to address the origin of the lepton mixing parameters of $U_\mathsf{MNSP}$ is with the implementation of a discrete flavor symmetry.  In this framework, a given mixing pattern is related to residual symmetries of the leptonic mass matrices which may arise from the spontaneous breaking of the flavor symmetry group. Simple models that utilize a spontaneously broken discrete flavor symmetry often predict zero leading order reactor mixing angle  and a maximal atmospheric mixing angle in the basis where the charged lepton mass matrix is diagonal.   
Examples include models in which the leading order neutrino mixing matrix $U_\nu$ described by tribimaximal (TBM) mixing~\cite{tbm1, tbm2, tbm3, tbm4,tbm5}, bimaximal (BM) mixing~\cite{bm1, bm2, bm3, bm4, bm5},  the two golden ratio mixing schemes (GR1~\cite{gr1} and  GR2~\cite{gr2}), or hexagonal (HEX)~\cite{hex} mixing.  Perhaps the simplest way to accommodate a nonzero reactor mixing angle while still using these popular starting points is to introduce a nontrivial lepton mixing matrix which can rotate $U_\mathsf{MNSP}=U_e^\dagger U_\nu$ away from its ``problematic'' leading order predictions. 

The emergence of a third nonzero lepton mixing angle gives rise to the possible appearance of the Dirac CP-violating phase originating in the charged lepton mixing matrix $U_e$ (assuming a particular form for $U_{\nu}$ as described above).  Together with the initial fixed starting point dictated by $U_\nu$, e.g., TBM or GR1 mixing, and an assumed form of $U_e$, it is possible to explore the predictions for the Dirac CP-violating phase $\delta$ which have mixing angles consistent with the measured experimental data.  The simplest of such assumed forms for $U_e$ is just a single rotation in the $1-2$ or $1-3$ planes.\footnote{A single rotation in the $2-3$ plane will not generate a correction to the (zero) reactor mixing angle.}  These corrections lead to relations, known as sum rules, between the model parameters contained in $U_e$, the leading order mixing angle predictions in $U_\nu$, and the experimentally measured angles in $U_\mathsf{MNSP}$.  One way to characterize the mixing angle predictions resulting from these sum rules has been to classify them into two types: atmospheric sum rules~\cite{atm2,atm1, atm3,atm4,atm5,atm6,atm7,atm8,atm9} and solar sum rules~\cite{sol1,sol2,sol3,sol4}.   While atmospheric sum rules arise from a variety of scenarios, solar sum rules are characteristic of models in which the leading order $U_{\nu}$ matrix is corrected by charged lepton contributions ~\cite{sum1,sum2a,sum2b,sum2,sum2c} (see also~\cite{sum3,sum3a, sum4,sum5,sum6,sum7,sum7a,sum8,sum9,sum10,sum11, sum12}, as well as the related work of~\cite{geref1,geref2,geref3,geref4,geref5,sumrulesDERS, sum13, Garg:2018, asymmetric}).
For a set of such models that involve a $1-2$ charged lepton rotation, there is a well-known sum rule for $\cos\delta$~\cite{sum3} (see also~\cite{sum2a,sum2b,sum2c, sum2,sum4,sum5,sumrulesDERS}):
\begin{equation}
\label{eq:sumruleorig}
\cos\delta=\frac{t_{23}s_{12}^2+s_{13}^2c_{12}^2/t_{23}-(s_{12}^{\nu})^
2(t_{23}+s_{13}^2/t_{23})}{s^\prime_{12}s_{13}},
\end{equation}
in which $c_{ij}=\cos\theta_{ij}$,
$s_{ij}=\sin\theta_{ij}$, $t_{ij} = \tan\theta_{ij}$, and primed letters represent the corresponding trigonometric functions of twice the argument, e.g., $s'_{ij}=\sin(2\theta_{ij})$. 
 As discussed extensively in the literature, the form of this sum rule (and the analogous sum rule for models with a $1-3$ charged lepton rotation) is quite striking in that it depends on just one model parameter, $(s_{12}^{\nu})^2$, and functions of the three observable mixing angles. This in turn has led to numerous studies of the posterior probability density function of $\cos\delta$ for a given $(s^\nu_{12})^2$, using the results of global fit data for the distributions of the lepton mixing angles as inputs. These results can provide guidance as to the precision needed in the determination of $\delta$ at current and forthcoming neutrino experiments to provide some degree of discriminating power in the theory space of possible models of lepton mixing.

However, in some cases, the bounded nature of the input parameters does not allow for the full range of experimentally allowed values for the lepton mixing angles to be accessed, such that care is needed to ensure that the posterior probability distribution of $\cos\delta$ falls fully in the physically allowed range, as needed to identify $\delta$ as a CP-violating phase. Here we explore this issue within a set of models that involve a single $1-2$ charged lepton rotation and satisfy Eq.~(\ref{eq:sumruleorig}). The scenarios considered include the cases mentioned above (BM, TBM, GR1, GR2, HEX), as well a new case with an unperturbed solar mixing angle $\pi/7$  (labeled here as NEW7), which we will introduce and discuss in the next section. We assume no non-standard interactions, such that the effects of additional field content such as flavons and other exotics are encoded by the given model parameters.  In these models, there are three continuously varying, bounded input parameters, including a single source of CP violation.  The predicted ranges of the observable mixing angles are thus restricted; these restrictions encode the needed constraints from the unitarity constraints of the lepton mixing matrix to ensure that $\delta$ is indeed to be identified as a physical CP-violating phase angle.\footnote{This includes for example the case of bimaximal mixing scenarios, which are known to be particularly constrained in terms of their ability to reproduce the preferred values of the lepton mixing angles as deduced from global fit data.} 

This paper is structured as follows. In Section~\ref{sec:background}, we describe the class of models considered and set up our approach.  The results are shown in Section~\ref{sec:approach2}. We summarize and conclude in Section~\ref{sec:conclusions}. Throughout this work, we use the global fit results from neutrino oscillation experiments as reported by the NuFIT collaboration~\cite{nufitweb} and summarized in Table~\ref{tab:1}. Indeed, we will see that the great precision of the measurement of the reactor angle $s^2_{13}$ significantly simplifies the analysis and provides nontrivial constraints on the feasibility of this set of models in predicting distributions for the remaining mixing parameters that are in reasonably good agreement with the data.
\begin{table}[H]
\centering
\begin{tabular}{|c|c|c|}
\hline 
 &  $3\sigma$ range \textbf{NO} &  $3 \sigma$ range \textbf{IO}\\
\hline 
$s^{2}_{12}$ & 0.275 $-$ 0.350 &0.275 $    -$ 0.350\\
$s^{2}_{23}$ & 0.427 $-$ 0.609 & 0.430 $   -$ 0.612\\
$s^{2}_{13}$ & 0.02046 $-$ 0.02440 & 0.02066 $-$ 0.02461\\
\hline \hline
\end{tabular}
\caption{The current status of the lepton mixing angles for the case of normal ordering (NO) and inverted ordering (IO), as taken from the global fit of~\cite{nufitweb}.}
\label{tab:1}
\end{table}


\section{Theoretical Background}
\label{sec:background}

We consider for concreteness a well-known class of theoretical models in which the starting point is the assumption that the matrix that diagonalizes the neutrino mass matrix is of the form
$U_{\nu}=R_{23}(\theta_{23}^{\nu})R_{12}(\theta_{12}^{\nu})$, where the $R_{ij}$ matrices are given by
\begin{equation}
\label{eq:neumix}
\begin{aligned}
R_{23}^{\nu}=\left(
\begin{array}{ccc}
 1 & 0 & 0 \\
 0 & c^{\nu}_{23} &  s^{\nu}_{23} \\
 0 & - s^{\nu}_{23} & c^{\nu}_{23} \\
\end{array}
\right), \;\;\; R_{12}^{\nu}=\left(
\begin{array}{ccc}
  c^{\nu}_{12} &  s^{\nu}_{12}&0 \\
  - s^{\nu}_{12} & c^{\nu}_{12}&0 \\
 0 & 0&1
\end{array}
\right),
\end{aligned}
\end{equation} 
and it is assumed that to leading order, the charged lepton mass matrix is diagonal in family space.  As a result, at leading order $s^2_{23}$ and $s^2_{12}$ are nonzero, while $s^2_{13}$ is zero.  The required shift to the reactor angle arises from corrections to the charged lepton sector, which here are encoded by a diagonalization matrix of the left-handed charged leptons in the $1-2$ plane.  

More precisely, the charged lepton diagonalization matrix is assumed to be of the form $U_{e}=U^e_{12}$, in which the $U^e_{ij}$ are defined as
\begin{equation}\label{rotations}
\begin{aligned}
	U_{23}^e & =\left(
\begin{array}{ccc}
 1 & 0        &                                0 \\
 0 & c^e_{23} &  s^e_{23}e^{-i \delta _{23}^{e}} \\
 0 & - s^e_{23}e^{i \delta _{23}^{e}} & c^e_{23} \\
\end{array}
\right), \;\;\; U_{12}^e=\left(
\begin{array}{ccc}
 c^e_{12} &  s^e_{12}e^{-i \delta _{12}^{e}}&0 \\
 - s^e_{12}e^{i \delta _{12}^{e}} & c^e_{12}&0 \\
 0 & 0&1
\end{array}
\right),\\
& \;\;\;\;\;\;\;\;\;\;\;\;\;\;\;\;\;\;\;\;\;\;\;
U_{13}^e  =\left(
 \begin{array}{ccc}
   c^e_{13} & 0& s^e_{13}e^{-i \delta _{13}^{e}}\\
   0 & 1&0\\
   - s^e_{13}e^{i \delta _{13}^{e}}&0 & c^e_{13} 
 \end{array}
 \right),
\end{aligned}
\end{equation}
in which $s^e_{ij} = \sin\theta^e_{ij}$ and $c^e_{ij} =
\cos\theta^e_{ij}$.\footnote{Note that there is an an intrinsic degeneracy in these definitions as 
$\delta_{ij}^{\prime e}\rightarrow \delta_{ij}^{e}-\pi$ and
$\theta_{ij}^{\prime e}\rightarrow \theta_{ij}^{e}-\pi /2$ yield the same
rotation matrix.  This will be taken into account in our analysis.}  Therefore, in models considered here, we have
\begin{equation}
U_\textsf{MNSP}\equiv U = U_e^\dagger U_\nu =U_{12}^{e\dagger}R_{23}^{\nu}R_{12}^{\nu}.
\label{eq:mnsp1}
\end{equation}
From this form of the lepton mixing matrix, the observable mixing angles take the form
\begin{equation}\label{eq:sij2}
\begin{aligned}
	s^2_{13} &=(s^e_{12})^2 (s^{\nu }_{23})^2,\;\;\;\;\; s^2_{23} =\frac{(s^{\nu }_{23})^2-(s^e_{12})^2 (s^{\nu }_{23})^2}{1-(s^e_{12})^2
		(s^{\nu }_{23})^2},\\
		s^2_{12} &=\frac{(c^{\nu }_{12})^2 (c^{\nu }_{23})^2 (s^e_{12})^2+(c^e_{12})^2 (s^{\nu }_{12})^2-2 c^e_{12} c^{\nu }_{12} c^{\nu }_{23}  s^e_{12} s^{\nu }_{12}\cos   \delta^e_{12}}{1-(s^e_{12})^2 (s^{\nu }_{23})^2},
\end{aligned}
\end{equation}
and the value of $\cos\delta$ is given by Eq.~(\ref{eq:sumruleorig}), subject to the constraints of Eqs.~(\ref{eq:sij2}).

Our goal is to explore these theoretical constraints among the observables within this well-studied class of models and their implications for the distribution of theoretically allowed values of $\cos\delta$.  The analysis is predicated on the nontrivial assumption that such models can provide a correct description of the observable lepton mixing parameters, together with specific regarding the probability distributions of the model parameters, as might emerge, {\it e.g.}, from a landscape picture or other possible ultraviolet completions of these models, though we remain agnostic as to their origin.  Here the model parameters are taken to be the following: $(s^e_{12})^2$, $(s^\nu_{23})^2$, $(s^\nu_{12})^2$, and $\cos\delta^e_{12}$.  The ``bare'' atmospheric neutrino mixing angle $\theta^\nu_{23}$ is often taken (or predicted) to be maximal, {\it i.e.}, that $(s^\nu_{23})^2=1/2$, but here  we will let it float.  This class of models can then be taken to be equivalent to the class of models with two rotations in the charged lepton sector, of the form $U_{e}=U_{23}^{e}(\theta_{23}^{e},\delta_{23}^{e})
U_{12}^{e}(\theta_{12}^{e},\delta_{12}^{e})$, with the phase $\delta^e_{23}=0$.  
Note that there is a single CP-violating phase, $\delta^e_{12}$, that sources the Dirac phase $\delta$.  As we will see, this feature yields a tight connection between the allowed values of $s^2_{12}$ and $\cos\delta$, which would clearly relax in situations with multiple CP-violating phases.  We defer the analysis of scenarios with multiple phases to future work.

While the parameters $(s^e_{12})^2$, $(s^\nu_{23})^2$, and $\cos\delta^e_{12}$ are continuous parameters, we consider only particular discrete values of $(s^\nu_{12})^2$ that correspond to values that can be achieved in specific models based on non-Abelian discrete family symmetries, as is standard in the literature.  As previously mentioned, these values correspond to the cases of bimaximal mixing (BM), tribimaximal mixing (TBM), hexagonal mixing (HEX), and two scenarios based on golden ratio mixing (GR1, GR2). We also introduce here a scenario we call the NEW7 scenario, in which the bare solar mixing angle is $\pi/7$. We note that
NEW7 can arise from preserving the Klein symmetry group\cite{bottomup2015}
\begin{equation} \nonumber
\begin{aligned} 
G_2=\left(
\begin{array}{ccc}
 -\sin \left(\frac{3 \pi }{14}\right) & \frac{\cos \left(\frac{3 \pi }{14}\right)}{\sqrt{2}} & \frac{\cos
   \left(\frac{3 \pi }{14}\right)}{\sqrt{2}} \\
 \frac{\cos \left(\frac{3 \pi }{14}\right)}{\sqrt{2}} & -\sin ^2\left(\frac{\pi }{7}\right) & \cos ^2\left(\frac{\pi
   }{7}\right) \\
 \frac{\cos \left(\frac{3 \pi }{14}\right)}{\sqrt{2}} & \cos ^2\left(\frac{\pi }{7}\right) & -\sin ^2\left(\frac{\pi
   }{7}\right) \\
\end{array}
\right),~G_3=\left(
\begin{array}{ccc}
 -1 & 0 & 0 \\
 0 & 0 & -1 \\
 0 & -1 & 0 \\
\end{array}
\right),~G_1=G_2G_3.
\end{aligned}
\end{equation}
Furthermore, it may also be possible to obtain NEW7 from a $D_{14}$ flavor symmetry group because a regular tetradecagon has an exterior angle of $\pi/7$.
Each of the models considered here thus have three continuous model parameters, and are broadly categorized by their specific value of $(s^\nu_{12})^2$, as given in Table~\ref{tab:2}.
\begin{table}[htb]
\centering
\begin{tabular}{|c|c|c|c|c|c|c|}
\hline 
 & BM & TBM  & HEX & GR1 & GR2 & NEW7 \\
\hline 
    $(s^\nu_{12})^2$ & $1/2$ & $1/3$ & $1/4$ & $(5-\sqrt{5})/10=0.276$ & $ (5-\sqrt{5})/8=0.345$ & $\sin^2(\pi/7)=0.188$ \\
\hline 
\end{tabular}
\caption{The values of $(s^\nu_{12})^2$
    for the theoretical scenarios
    under consideration.} 
\label{tab:2}
\end{table} 
We see that of these scenarios, the values of $(s^{\nu}_{12})^2$ are similar for TBM and GR2, and also for HEX and GR1, and for each, relatively small corrections are needed to be in the experimentally preferred range for the solar mixing angle, as given in Table~\ref{tab:1}.  In contrast, the BM and NEW7 scenarios both correspond to values of $(s^{\nu}_{12})^2$ that are well out of the preferred range, with BM larger and NEW7 smaller.  This feature of BM and NEW7 will require particular care when determining the posterior probability distribution for the Dirac lepton mixing phase.


\section{Analysis and Results}
\label{sec:approach2}

From this starting point, we now consider the possibilities for the probability distributions for the three continuous model parameters with the goal of predicting distributions for the observables  mixing angles in alignment with the global fit data of~\cite{nufitweb}, and determining the resulting probability distribution for $\cos\delta$.  
For notational simplicity, we relabel the model parameters as follows:
\begin{equation}
a\equiv\left(s_{12}^e\right)^2, \qquad 
b\equiv\left(s_{23}^\nu\right)^2, \qquad
c\equiv\cos\delta_{12}^e.
\end{equation}
We also relabel the observable mixing angles $s^2_{13}$, $s^2_{23}$, and $s^2_{12}$ as $x$, $y$, and $z$, respectively. Finally, we define $z_0\equiv (s^\nu_{12})^2$, as this quantity is the ``bare'' value of $s^2_{12}\equiv z$.  

\begin{figure}[tb]
    \begin{minipage}{0.5\textwidth}
        \centering
        \includegraphics[scale=\imgscale]{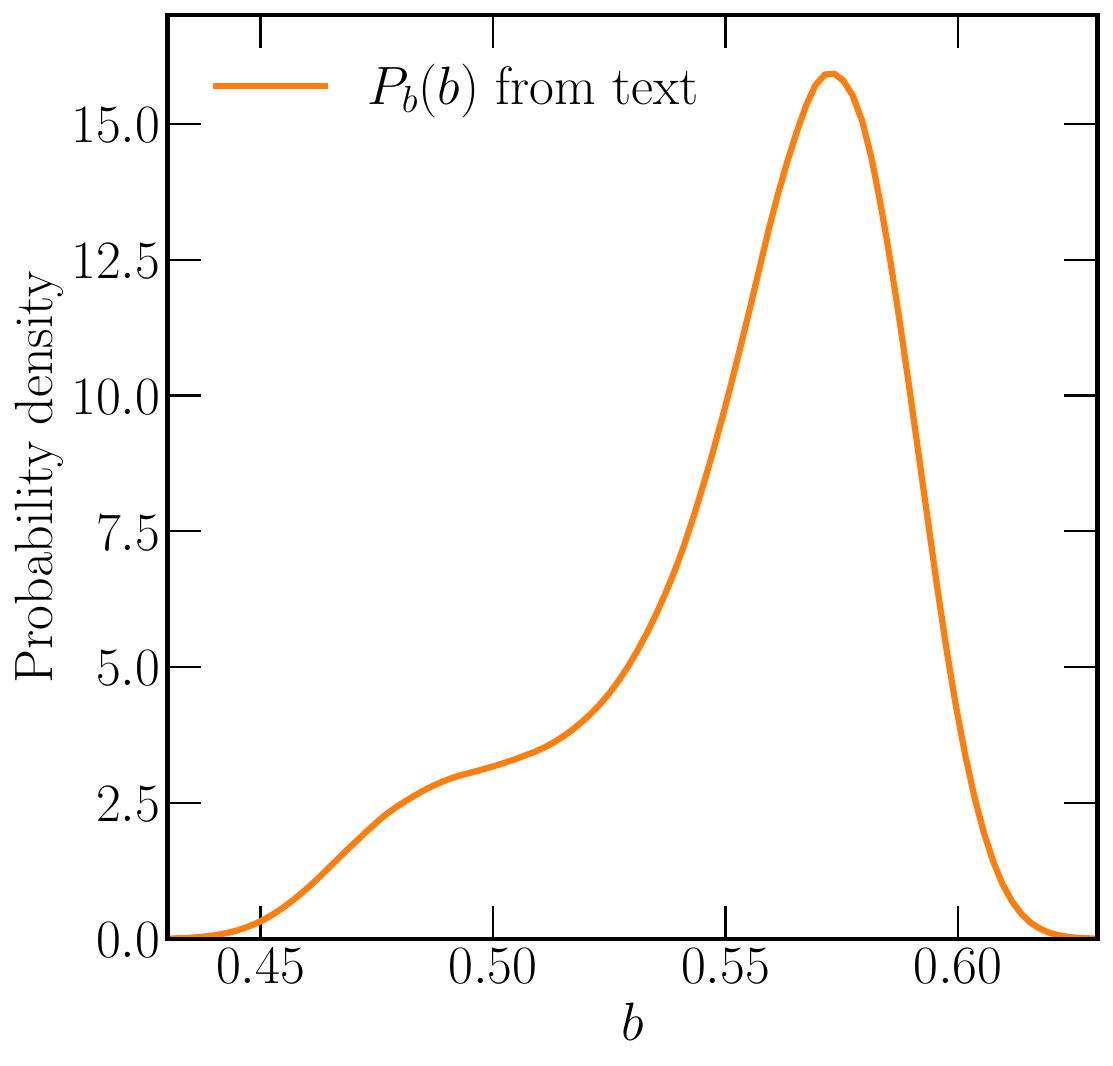}
    \end{minipage}
    \begin{minipage}{0.5\textwidth}
        \centering
        \includegraphics[scale=\imgscale]{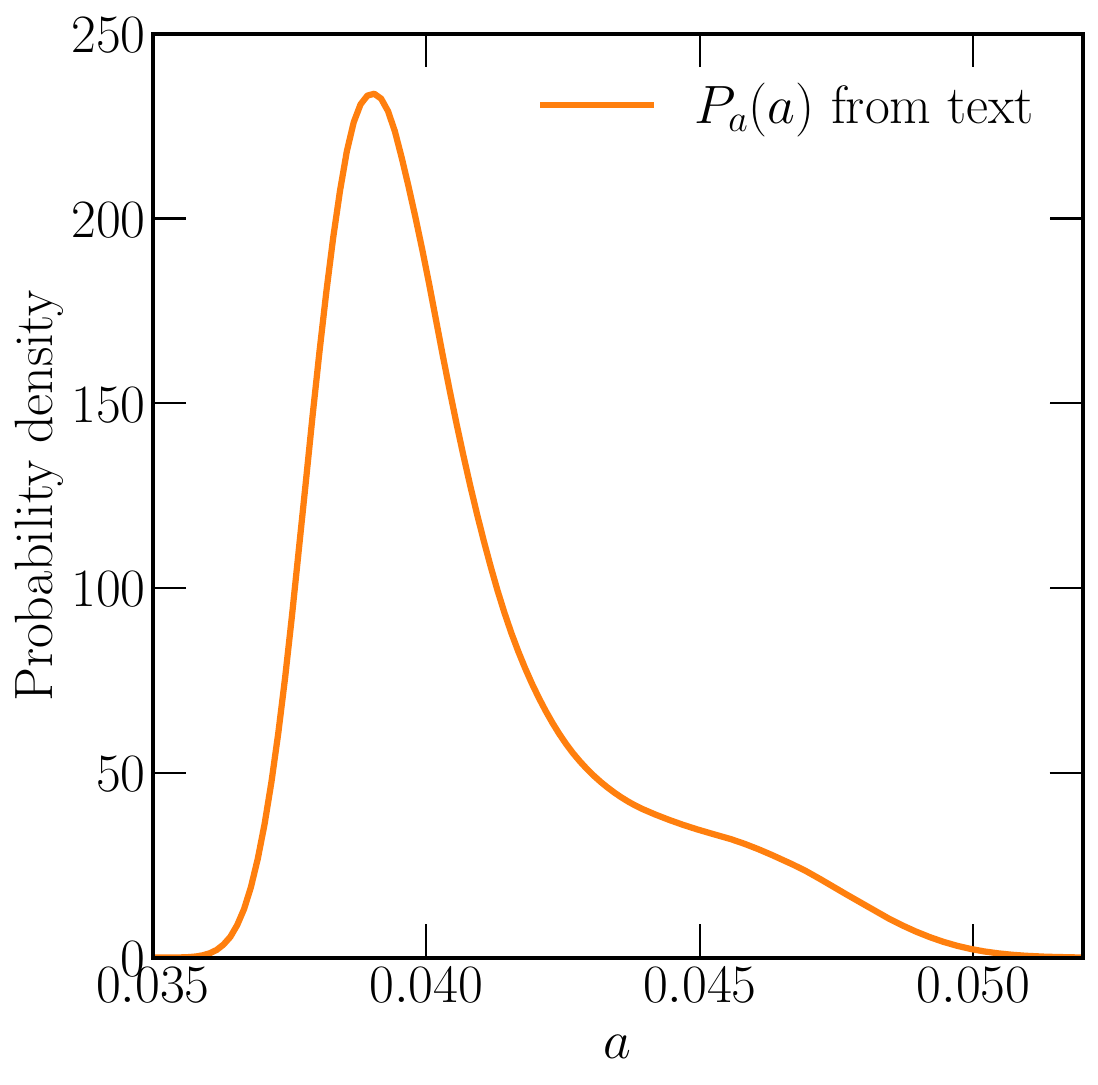}
    \end{minipage}
    \caption{
        Distributions for the model parameters $b$ (left) and $a$ (right) from
        Eqs.~\eqref{eq:PbPy} and~\eqref{eq:PaPy}.
        \label{fig:PaPb}
        }
\end{figure}

An immediate simplification results from the fact that the reactor angle has been measured with great precision.  As such, we can assume to leading order that the distribution for $x\equiv s^2_{13}$ can be effectively modeled as a delta function, fixed about its central value of $(s^2_{13})_0\equiv x_0=0.02241$~\cite{nufitweb}\footnote{Here we take the central value of $s^2_{13}$ for the case of normal ordering, for concreteness.}, as follows:
\begin{equation}
P_x(x)=\delta(x-x_0).
\label{eq:pxdistribution}
\end{equation}  
With this in hand, we begin with the atmospheric mixing angle distribution, which we label as $P_y(y)$.  From Eq.~(\ref{eq:sij2}), we see that both $x$ and $y$ only depend on the model parameters $a$ and $b$. 
As we consider here the case that the distribution of $x$ is fixed to its central value as in Eq.~(\ref{eq:pxdistribution}), we have
\begin{align}
\label{eq:xabx0}
x = ab = x_0\:\:\: \Rightarrow\:\:\: a = \frac{x_0}{b}\:\:\: \text{or}\:\:\: b
= \frac{x_0}{a}.
\end{align}
Taking some freedom in the choice of notation, let us represent as
$P_{\alpha|\beta}(\alpha)$ the distribution of some variable $\alpha$ for a
given value of the variable $\beta$, we can then write
\begin{align}
\label{eq:PabPba}
P_{a|b}(a) = \deltafun\left(a - \frac{x_0}{b}\right)\:\:\:\: \text{and}\:\:\:\:
P_{b|a}(b) = \deltafun\left(b - \frac{x_0}{a}\right).
\end{align}
Note that, in general, $P_{\alpha|\beta}(\alpha)$ distributions are not
equivalent to $P_\alpha(\alpha)$, which will represent the marginalized
distribution of the variable $\alpha$.
Only in cases where $P_{\alpha|\beta}(\alpha)$ is not a function of
$\beta$ we can say $P_{\alpha|\beta}(\alpha) = P_\alpha(\alpha)$.
Assuming $P_b(b)$ can be integrated to unity, it is straightforward to show that
\begin{equation}
\label{eq:Pxint}
P_x(x) = \int da\, db\, \deltafun\left(ab - x\right) P_{a|b}(a) P_b(b) = \deltafun\left(x - x_0\right),
\end{equation}
which should have the same result had we started with $P_{b|a}(b) P_a(a)$
and assumed instead that $P_a(a)$ is a normalized probability density.
So far, Eq.~\eqref{eq:Pxint} serves mostly as a check since it returns the
same distribution that we used to obtain Eq.~\eqref{eq:PabPba}. For the $P_y(y)$ integral, 
is easy to check that from Eq.~\eqref{eq:PabPba}
\begin{align}
\label{eq:Pyint}
P_y(y) & = \int da\, db\, \deltafun\left(\frac{b - ab}{1 - ab} - y\right) P_{a|b}(a) P_b(b) 
	= (1 - x_0) P_b(y + x_0(1 - y)).
\end{align}
We can use this result to solve for the marginalized distribution
of $b$, $P_b(b)$, as
\begin{align}
\label{eq:PbPy}
P_b(b) = \frac{1}{1 - x_0}\, P_y\!\left(\frac{b - x_0}{1 - x_0}\right).
\end{align}
If we instead start with $P_{b|a}(b) P_a(a)$
and follow a similar approach we can find the
marginalized distribution $P_a(a)$ given by
\begin{align}
\label{eq:PaPy}
P_a(a) = \frac{x_0}{a^2(1 - x_0)}\, P_y\!\left(\frac{x_0(1 - a)}{a(1 - x_0)}\right).
\end{align}
In Figure~\ref{fig:PaPb}, the resulting probability distributions from
Eqs.~\eqref{eq:PbPy} and~\eqref{eq:PaPy} are shown as orange solid lines.  Note that if the distributions of $a$ and $b$ were assumed to be independent, $P_y(y)$ would be predicted to be a delta function.

The integral for $P_z(z)$ depends on the distributions of all three model parameters $a$,
$b$ and $c$, so that at least two of the parameter distributions are to be conditioned on another parameter.
Consider $P_{a|b}(a) P_{b|c}(b) P_c(c)$. Since we do not know $P_{b|c}(b)$ and $z(a=x_0/b, b, c)$ can only give upper and lower limits for b depending on $c$, $z$ and $z_0$, let us
assume that $b$ is independent of $c$ and $P_{b|c}(b) = P_b(b)$, where
$P_b(b)$ is given by Eq.~\eqref{eq:PbPy}.
Therefore, we have 
\begin{align}
\label{eq:Pzintstart}
P_z(z) = \int da\, db\, dc\, \deltafun\left(f - z\right) P_{a|b}(a) P_b(b) P_c(c),
\end{align}
where $f=z_0 - c d_1 + d_2$, with 
\begin{align}
d_1 & = 2\frac{\sqrt{(1 - b)(b - ab)(1 - z_0)a b z_0}}{b(1 -ab)}, \qquad d_2  = \frac{a b(1 - b)(1 - 2z_0)}{b(1 - ab)}.\nonumber
\end{align}
We note that since $a$, $b$ and $z_0$ are always between 0 and 1, $d_1$ and
$d_2$ are always positive.
Eq.~\eqref{eq:Pzintstart} can then be written as
\begin{align}
P_z(z) & = \int db\, dc\, \deltafun\left(\eval[2]{f}_{a=\frac{x_0}{b}} - z\right) P_b(b) P_c(c) \nonumber\\
       \label{eq:Pzint}
       & = \int db\, \frac{b(1 - x_0)}{2\sqrt{(1 - b)(b - x_0)(1 - z_0)x_0
               z_0}}\,  P_b(b) P_c(c_0(b, z)),
\end{align}
where 
\begin{align}
\label{eq:thec0}
c_0(b, z) = \frac{b(x_0 - 1)(z - z_0) + x_0(1 - b)(1 - 2z_0)}
{2\sqrt{x_0 z_0(1 - b)(b - x_0)(-z_0 + 1)}}.
\end{align}
We can try to guess the shape of $P_c(c_0(b,z))$ as a curve that depends
on a few parameters and find the best values for said parameters by doing
the integral for a few values of $z$. For example, if $P_c$ is taken to be a normalized Gaussian curve, we would
need to evaluate the integral for at least two different values of $z$
to obtain an estimation of the center of the peak (the mean of $c$) and the
width of the distribution (related to the standard deviation of $c$).  As discussed in the Appendix, we can also choose modified distributions such as the skew normal distribution $P_{\rm skew}$,  or the Gram-Charlier distribution $P_{\rm GC}$.  Any of these distributions can be used as $P_c(c)$ in Eq.~\eqref{eq:Pzint} (both $P_{\rm skew}$ and $P_{\rm GC}$ reduce to $P_{\rm Gauss}$ for
$s=k=0$). To properly estimate $P_c(c)$ we need to use at least as many test $z$ values
as parameters in the choice of distribution functions.

The procedure is then: (i) select one of the density functions
to use as $P_c$ in Eq.~\eqref{eq:Pzint}, (ii) select a series of test
$z$ values, (iii) evaluate the integral for each test $z$ value, (iv) compare
the estimation from the integral with the actual value of $P_z(z)$ from global
fit, (v) minimize the difference between actual and estimated $P_z(z)$ by
changing the density function parameters.
The fourth step could be carried out using absolute differences between
estimated and actual $P_z(z)$ or weighted with the actual value of $P_z(z)$
from global fit to give more relevance to points with higher probability.
Here the test points that will be used are the central value of
$z$, the $\pm 1\sigma$ values, the $\pm 3\sigma$ values and the points at half
distance between $\pm 1\sigma$ and $\pm 3\sigma$ which, presumably, should be
close to $\pm 2\sigma$.
Weighted differences between the actual and estimated $P_z(z)$ will be used so
that in cases where the probability density function can only give good
precision for a few points, the points with higher probability are selected.

The best parameter values for each density function in
Eqs.~\eqref{eq:gaussnorm}--\eqref{eq:gaussgc} to work as $P_c(c_0(b, z))$ in
the integral of Eq.~\eqref{eq:Pzint} are given in
Table~\ref{table:densityparameters} for four neutrino mixing patterns.
Using $P_{\rm skew}$ to estimate $P_z(z)$ does not show a significant deviation
from $P_{\rm Gauss}$ since the skewness is small ($s\approx 10^{-2}$).
However, $P_{\rm CG}$ shows a larger deviation mostly for
TBM and GR2 on the skewness ($s$) side, and for GR2 and HEX in kurtosis ($k$).
After finding the best parameters for each of the three density functions, we
can further compare several points of the actual (from the global fit) and
estimated $P_z(z)$ (from Eq.~\eqref{eq:Pzint}) to determine which density
function gives the best results for each neutrino mixing pattern.

\begin{table}[tb]
\begin{center}
\begin{tabular}{|c|ccc|}
\hline
    & $P_{\rm Gauss}$ & $P_{\rm skew}$ & $P_{\rm CG}$ \\
    & $(\mu,\sigma)$ & $(\mu,\sigma,s)$ & $(\mu,\sigma,s,k)$ \\
\hline
TBM & $( 0.233,\, 0.0981)$ & $( 0.233,\, 0.0982,\, 0.0106)$ & $( 0.228,\, 0.0974,\, -0.0894,\, -0.0885)$\\
HEX  & $(-0.440,\, 0.10  )$ & $(-0.441,\, 0.10  ,\, 0.0145)$ & $(-0.439,\, 0.0982,\,  0.0290,\, -0.199 )$\\
GR1 & $(-0.213,\, 0.101 )$ & $(-0.214,\, 0.101 ,\, 0.0142)$ & $(-0.214,\, 0.10  ,\, -0.0235,\, -0.0792)$\\
GR2 & $( 0.324,\, 0.0966)$ & $( 0.323,\, 0.0966,\, 0.0102)$ & $( 0.317,\, 0.0956,\, -0.106 ,\, -0.121 )$\\
\hline
\end{tabular}
\end{center}
\caption{
	Best parameter values for each probability density function to work as
	$P_c(c_0(b, z))$ in the integral of Eq.~\eqref{eq:Pzint} for the TBM, HEX, GR1, and GR2 mixing patterns.
	\label{table:densityparameters}
}
\end{table}

\begin{figure}[tb]
    \begin{minipage}{0.5\textwidth}
        \centering
        \includegraphics[scale=\imgscale]{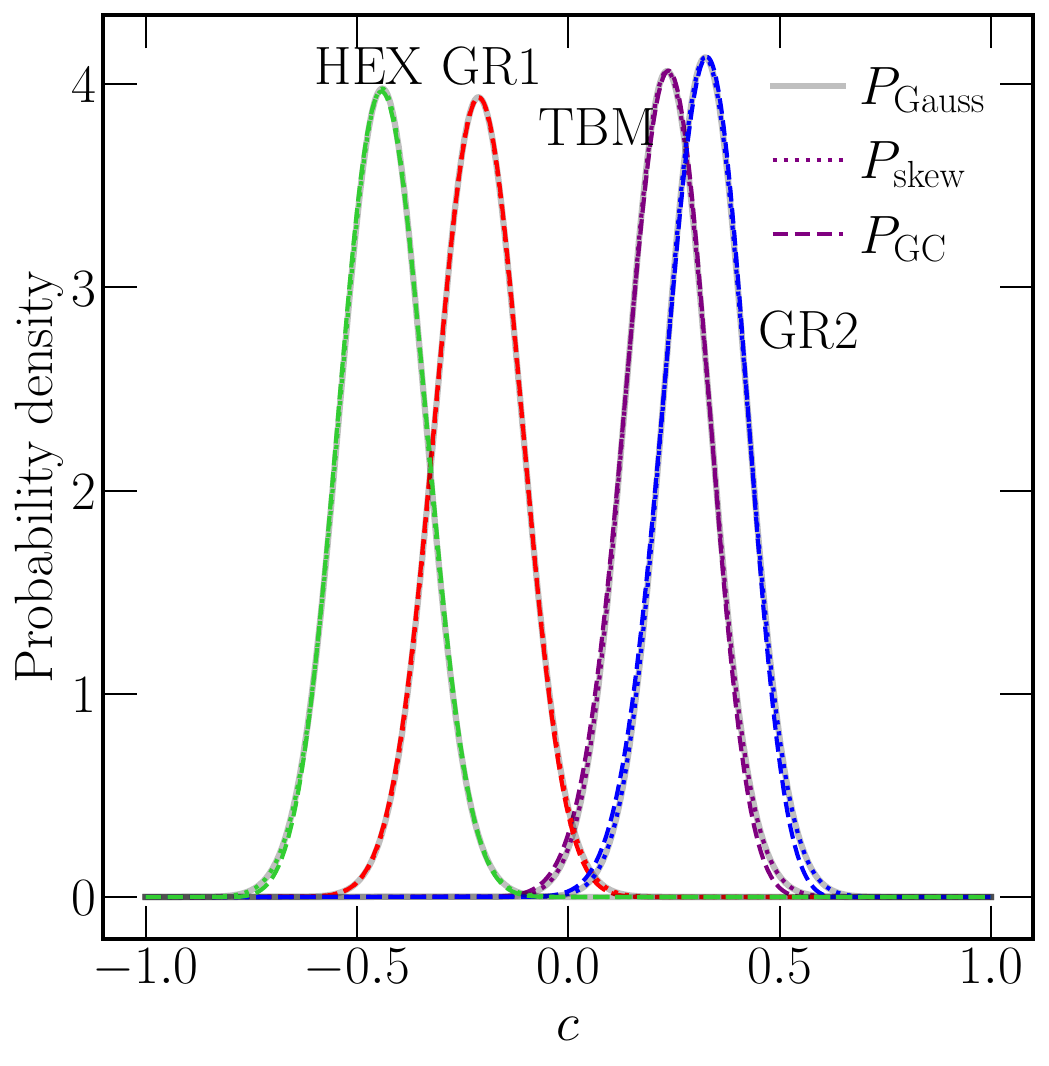}
    \end{minipage}%
    \begin{minipage}{0.5\textwidth}
        \centering
        \includegraphics[scale=\imgscale]{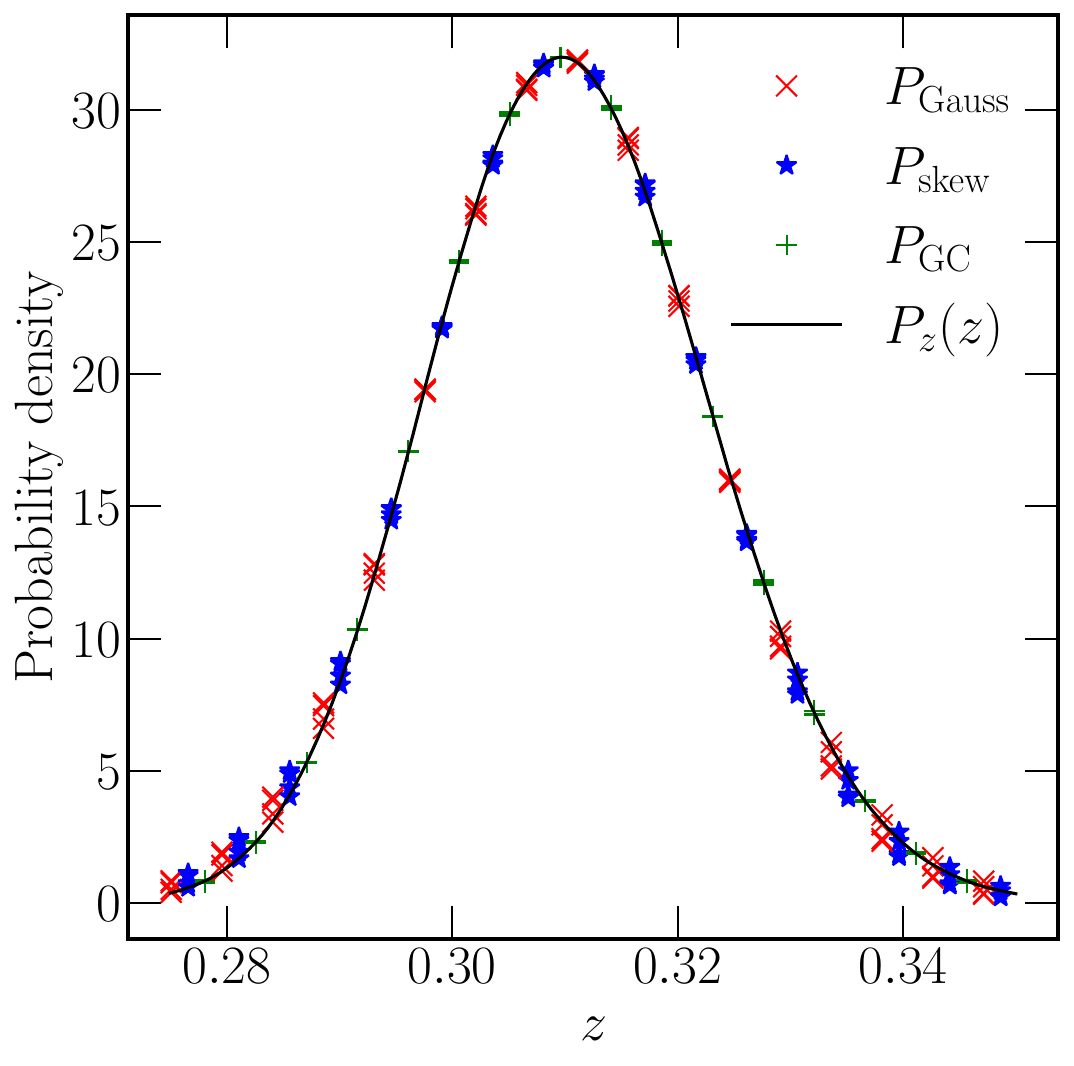}
    \end{minipage}
    \caption{
        Comparisons between the different choices used for $P_c$.
        On the left, the resulting $P_c(c)$
        is shown for TBM, HEX, GR1, and GR2,
        and the choices $P_{\rm Gauss}$ (gray solid),
        $P_{\rm skew}$ (dotted colored) and $P_{\rm GC}$ (dashed colored).
        On the right, a comparison between estimated and actual $P_z(z)$
        for the three different density function choices
        showing different degrees of spreading
        due to the different mixing patterns.
        \label{fig:guesscompare}
        }
\end{figure}

The density function that best estimates $P_z(z)$ is $P_{GC}$, as expected since this distribution involves more parameters.  The results are shown in Figure~\ref{fig:guesscompare}.
On the right we can see that $P_{\rm skew}$ follows $P_{\rm Gauss}$ very
closely, while $P_{\rm GC}$ shows some deviation from the others for densities
below 1.
On the right, the resulting estimation for $P_z(z)$ inside the $\pm 3\sigma$
range is compared with the actual values from the global fit.
$P_{\rm Gauss}$ (red crosses) and $P_{\rm skew}$ (blue stars) approximate the
global fit based probability density mostly at the top of the distribution but
become more inaccurate for densities below 10.
Both present a large amount of spreading between different
patterns as shown for points with the same $z$ value but different density
values. However,
the values from $P_{\rm GC}$ match the global fit based curve
with very good accuracy in the whole $\pm 3\sigma$ range of $z$
with little to no spreading between different mixing patterns.
Hence, the $P_c(c)$ distributions will be based on $P_{\rm GC}$,
since this choice provides the best match for the global fit based $P_z(z)$.%
\footnote{
    We note here that while the observed symmetry of $P_c(c)$ in the
    left panel of Figure~\ref{fig:guesscompare} may seem nonintuitive given
    the symmetry of the experimental distribution of $P_z(z)$ and the
    asymmetric form of $P_b(b)$, this result emerges due to the nature of the
    conditional probability $P_c(c(b)_z)$ and the form of
    Eq.~(\ref{eq:Pzint}).}

\begin{figure}[tb]
	\centering
	\includegraphics[scale=\imgscale]{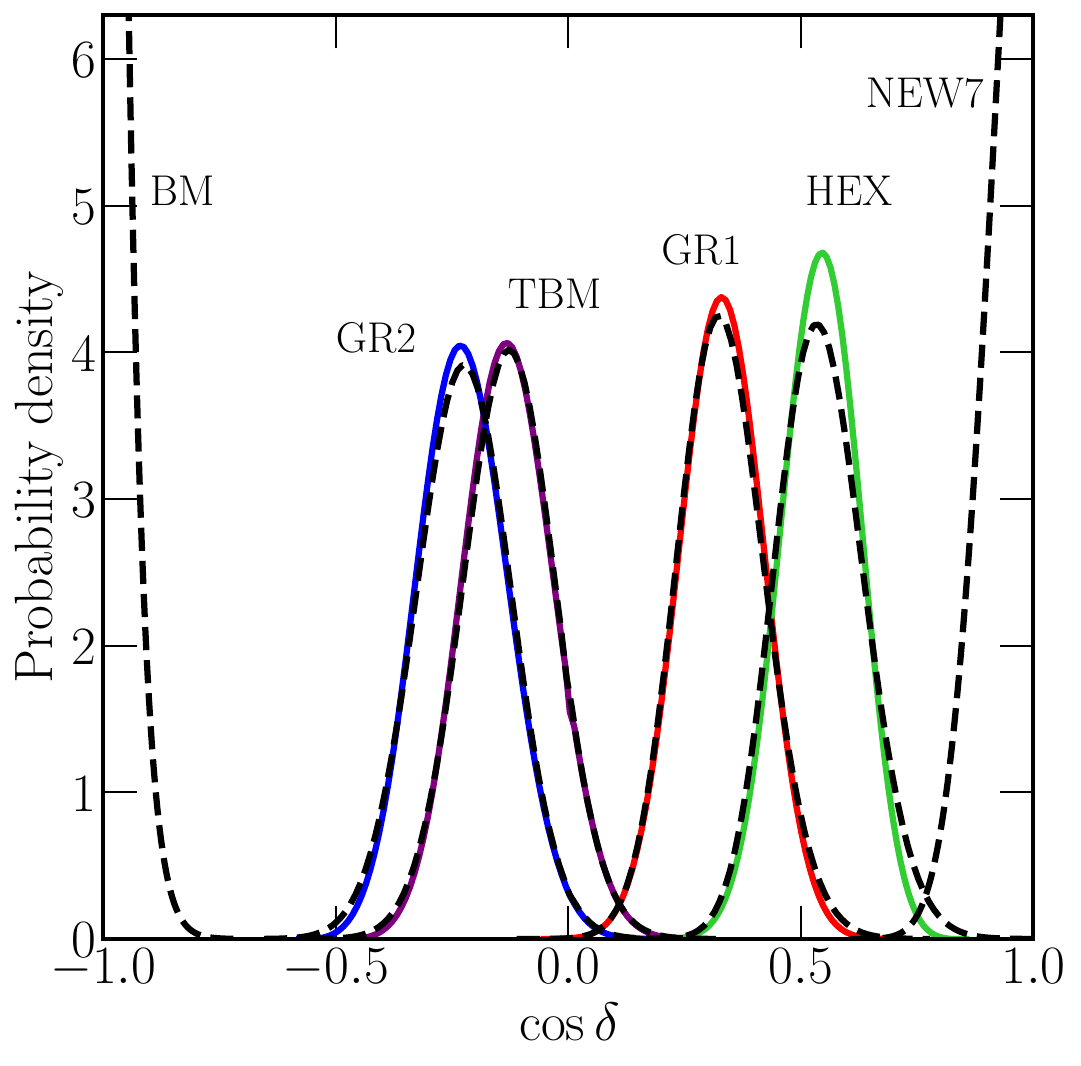}
	\caption{
		Probability densities for $\cos\delta$ from the integral in
		Eq.~\eqref{eq:Pcosintfinal} are shown in color.
		The dashed black lines were obtained from the integral in
		Eq.~\eqref{eq:PcosintfinalPbc}.
		\label{fig:Pcosdelta}
	}
\end{figure}

Given our assumption that $x$ follows a delta-function distribution we can
replace $x\to x_0$.
The integral for the probability density of $\cos\delta$ is  
\begin{align}
P_{\cos\delta}(\cos\delta) = \int da\, db\, dc\, \delta(\tilde{g} - \cos\delta) P_{a|b}(a) P_b(b) P_c(c),
\end{align}
where similarly to $P_z(z)$, it was assumed that $P_{b|c}(b) = P_b(b)$ and
$P_b(b)$ is given by Eq.~\eqref{eq:PbPy}. Here we have used the shorthand notation $\tilde{g}(x,b,z)$ to express $\cos\delta$ in terms of these parameters, as follows:
\begin{align}
\tilde{g} = 
	\frac{(b - x)z + x(1 - z)(1 - b) - z_0 b(1 - x)}
	    {2\sqrt{zx(1 - z)(1 - b)(b - x)}},
\end{align}
where $x=x(a,b)$ and $z=z(a,b,c)$ are functions of the model parameters. Thus, we have
\begin{align}
    P_{\cos\delta}(\cos\delta)
    & = \int da\, db\, dc\, \delta(\tilde{g} - \cos\delta)
    \delta\left(a - \frac{x_0}{b}\right) P_b(b) P_c(c)\nonumber\\
    \label{eq:Pcosintfinal}
    & = \int dc\,
    \eval{\left[\left(\frac{\partial \tilde{g}}{\partial b}\right)^{-1}
            P_b(b) P_c(c)\right]}_{b\text{ such that } \tilde{g} = \cos\delta},
\end{align}
The integrand in the last equation has to have $b$ values that give
$\tilde{g} = \cos\delta$ depending on the values of $c$.
In other words, when we integrate over $c$ the value of
$b$ will change such that $\tilde{g}(a=x_0/b, b, c) =
\cos\delta$.
Using $P_b(b)$ from Eq.~\eqref{eq:PbPy}
and $P_c(c)$ obtained with the parameters of Table~\ref{table:densityparameters}
we can evaluate Eq.~\eqref{eq:Pcosintfinal},
choosing $z_0$ values corresponding to the TBM, HEX, GR1 and GR2
neutrino mixing patterns.
The results
are shown as colored lines in Figure~\ref{fig:Pcosdelta}.

\begin{figure}[tb]
	\begin{minipage}{0.5\textwidth}
		\centering
		\includegraphics[scale=\imgscale]{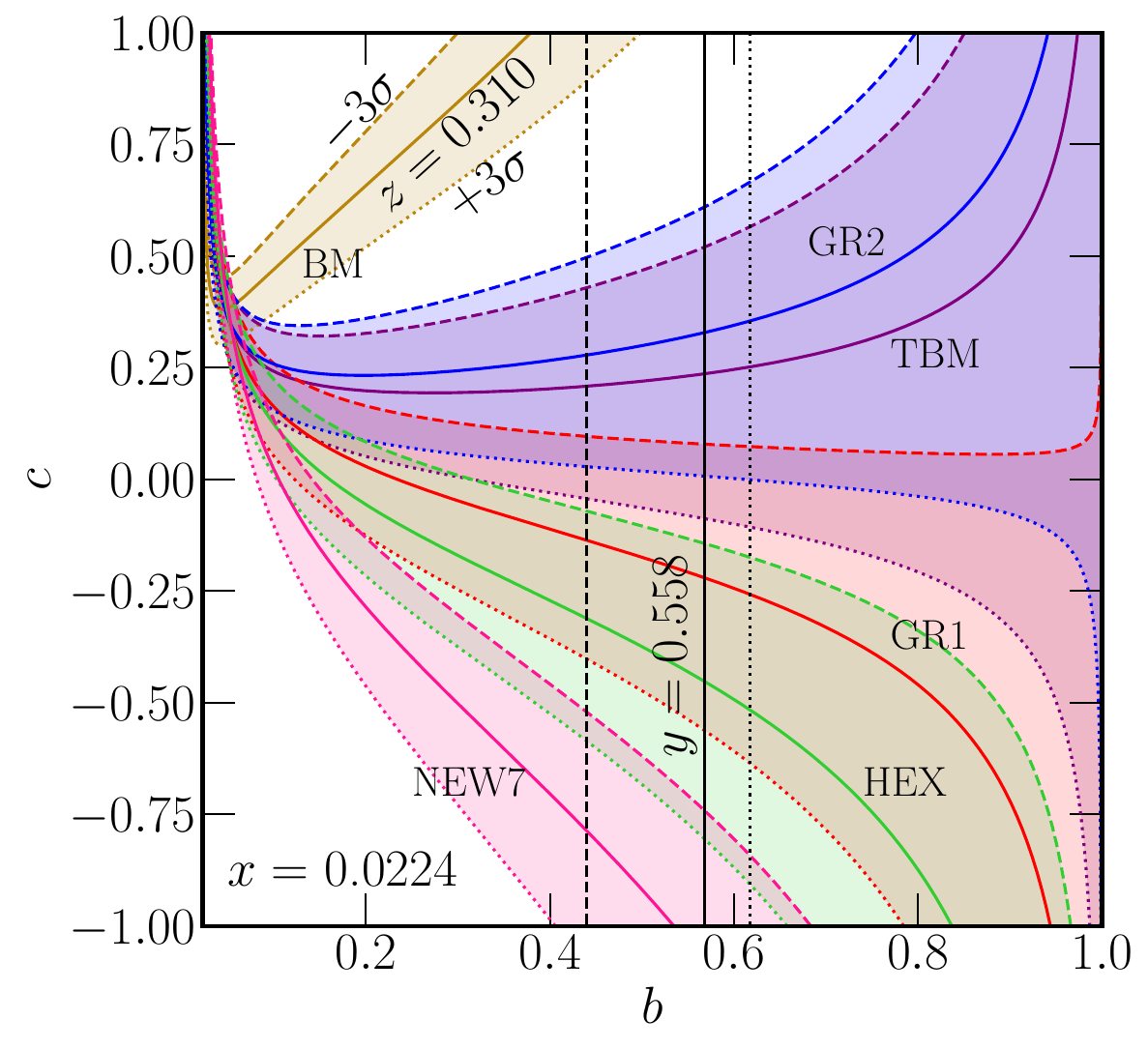}
	\end{minipage}%
	\begin{minipage}{0.5\textwidth}
		\centering
		\includegraphics[scale=\imgscale]{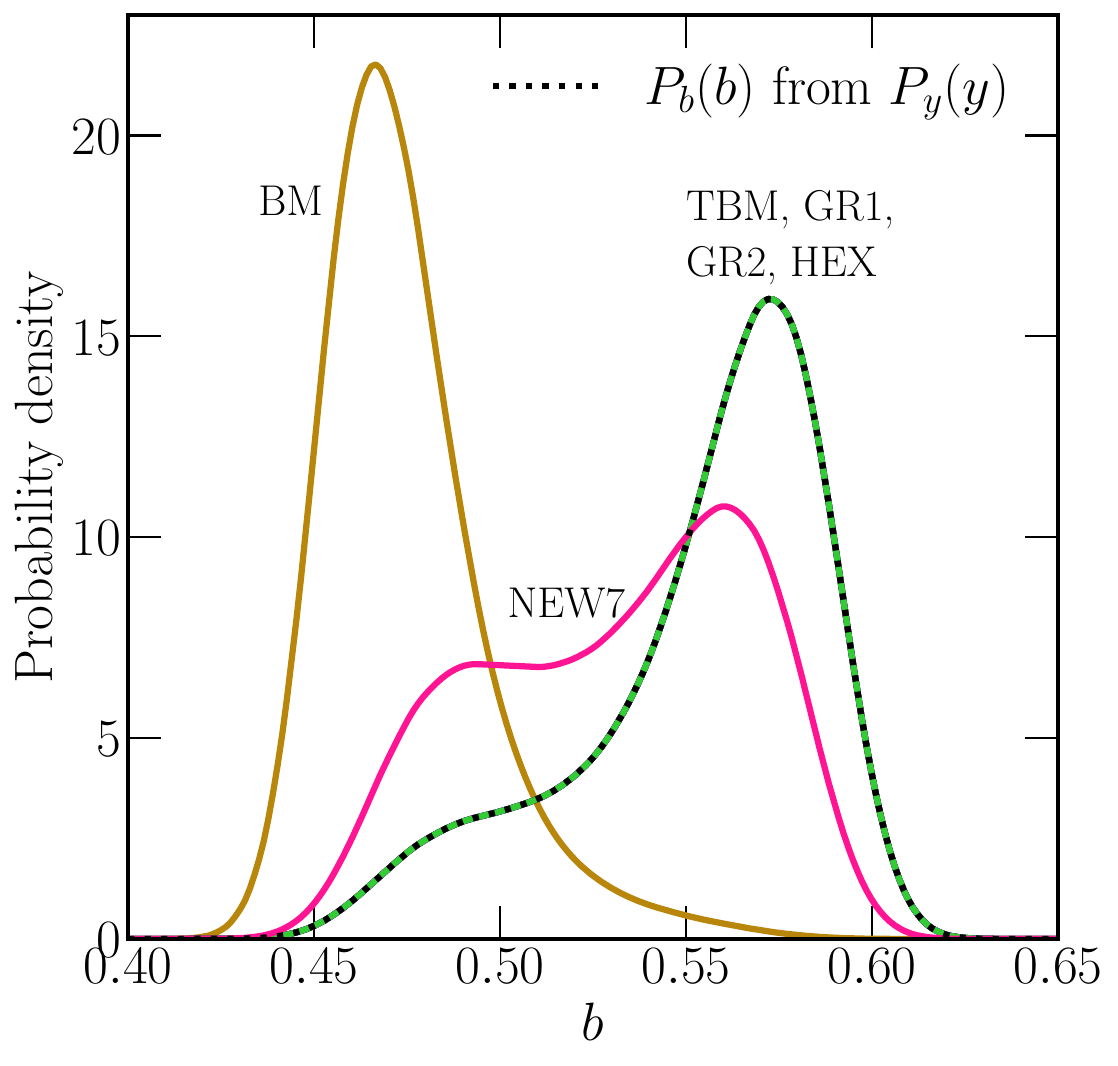}
	\end{minipage}
	\caption{
		Left: Lines for the central (solid), $-3\sigma$ (dashed) and
		$+3\sigma$ (dotted) values of $z=\sin^2\theta_{12}$ are shown in color for different
		patterns and in black for $y=\sin^2 \theta_{23}$.
		Right: The probability distributions obtained from marginalizing
		$c$ out of $P_{b,c}(b, c)$ for all the patterns.
		The dotted black line is the distribution $P_b(b)$ of
		Eq.~\eqref{eq:PbPy}.
		\label{fig:zlines}
	}
\end{figure}

Due to the experimental constraints, the BM and NEW7 patterns have certain novel features that require a more refined analysis. As seen in Figure~\ref{fig:zlines},
in these two cases the central values for $y$ and $z$
never meet inside the valid $b$-$c$ rectangle.
Moreover, the $z$-constant lines show
that $c$ has a stronger $b$-dependence,
rapidly growing to 1 and $-1$ for BM and NEW7, respectively.
The assumption $P_{b|c}(b) = P_b(b)$
is thus expected to break down for these two patterns.
For these cases, we will instead work with the two-dimensional probability $P_{b,c}(b, c)$
that can be obtained from the two dimensional $\chi^2$ projection
for $s^2_{12} = z$ and $s^2_{23} = y$ available at the NuFIT
website~\cite{nufitweb}. 
A density probability can be obtained by normalizing the likelihood obtained
from $\exp(-\chi^2(y,z)/2)$ such that the two-dimensional integral over the domain
of $y$ and $z$ is equal to 1.

\begin{table}[tb]
\begin{center}
\begin{tabular}{|c|c|}
\hline
	& factor ($\mathcal{N}$) \\
\hline
BM  & 1271.95 \\
TBM & $1.0 + \mathcal{O}(10^{-10})$ \\
HEX  & $1.0 + 9.67\times 10^{-6}$ \\
GR1 & $1.0 + \mathcal{O}(10^{-9})$ \\
GR2 & $1.0 + \mathcal{O}(10^{-10})$ \\
NEW7 & 2.80 \\
\hline
\end{tabular}
\end{center}
\caption{
	Factors needed to normalize the (pre)distribution $\bar{P}_{b,c}(b,c)$, as
	in Eq.~\eqref{eq:2ndnorm}, to obtain a proper probability density
	$P_{b,c}(b,c)$.
	\label{table:2ndnorm}
}
\end{table}

This joint distribution, which we will call $P_{y,z}(y,z)$, is related to the joint (pre)distribution
of $b$ and $c$, $\bar{P}_{b,c}(b,c)$, as follows:
\begin{align}
\label{eq:Pyzint}
P_{y,z}(y,z) & = \int db\, dc\, \deltafun\left(y(b) - y\right)
    \deltafun\left(z(b,c) - z\right) \bar{P}_{b,c}(b,c)\nonumber \\
             & = \frac{\left(1 - x_0\right)^{3/2} \left[1 + x_0 + y (1 - x_0)\right]}
             {2 \sqrt{x_0 y z_0 (1 - z_0)\left[1 - x_0 - y(1 - x_0)\right]}}
             \bar{P}_{b,c}\left(x_0 + y(1 - x_0),\frac{d_2 + z_0 -z}{d_1}\right),
\end{align}
where $y(b)$ and $z(b,c)$ are obtained after fixing $a=x_0/b$, and
the bar above $\bar{P}$ is related to the (pre)distribution name, as
explained later.
Solving for $\bar{P}_{b,c}(b,c)$, we find 
\begin{align}
\label{eq:Pbcbc}
\bar{P}_{b,c}(b,c) = \frac{2 \sqrt{x_0z_0(1 - b)(b - x_0)(1 - z_{0})}}
    {b \left(1 - x_0\right)^2} P_{y,z}(y(b), z(b,c)).
\end{align}
Recall that $P_{y,z}(y,z)$ was defined from the normalized likelihood, which
was obtained from the 2-dimensional $\chi^2(x, z)$, and should integrate to $1$ inside
the valid $y$-$z$ rectangle.
However, the integration of Eq.~\eqref{eq:Pyzint}
is done over the allowed $b$-$c$ rectangle
and therefore is valid only
for the combinations of $y(b)$ and $z(b,c)$
that can be obtained inside the model parameters' domain, which is
pattern dependent.
As a result, the (pre)distribution $\bar{P}_{b,c}(b,c)$ in
Eq.~\eqref{eq:Pbcbc} needs to be normalized once again in the domain of $b$ and $c$ to ensure a correct probability density.
The additional normalization
can be seen as a measure of how well
a particular pattern repeats the global fit values
of $y$ and $z$.
Patterns that easily fit inside the $\pm3\sigma$ global fit range
for both quantities, such as TBM and GR2
(see Figure~\ref{fig:zlines}, left side),
will have a normalization factor that differ from one by less than $10^{-10}$.
Other patterns, such as HEX, NEW7 and BM,
differ from one by nearly $10^{-5}$, 2.8 and more than $10^3$,
respectively.
\begin{equation}
\label{eq:2ndnorm}
P_{b,c}(b,c) = \mathcal{N}[\text{pattern}]\times\bar{P}_{b,c}(b,c)
\end{equation}
where $P_{b,c}(b,c)$ represents the proper probability density that can be
integrated to one in the domain of $b$ and $c$.
The required factors are shown in Table~\ref{table:2ndnorm}.

\begin{figure}[tb]
	\begin{minipage}{0.5\textwidth}
		\centering
		\includegraphics[scale=\imgscale]{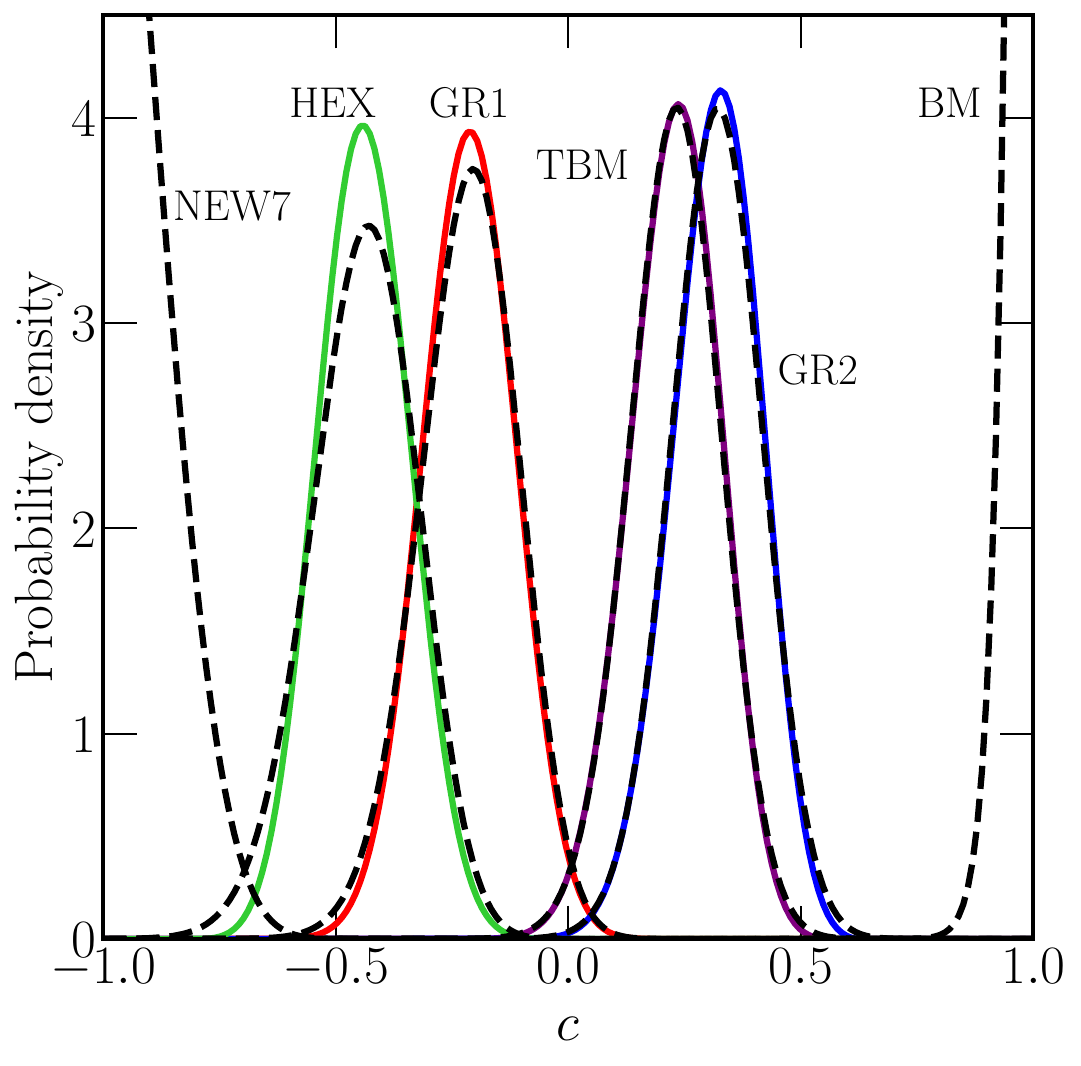}
	\end{minipage}%
	\begin{minipage}{0.5\textwidth}
		\centering
		\includegraphics[scale=\imgscale]{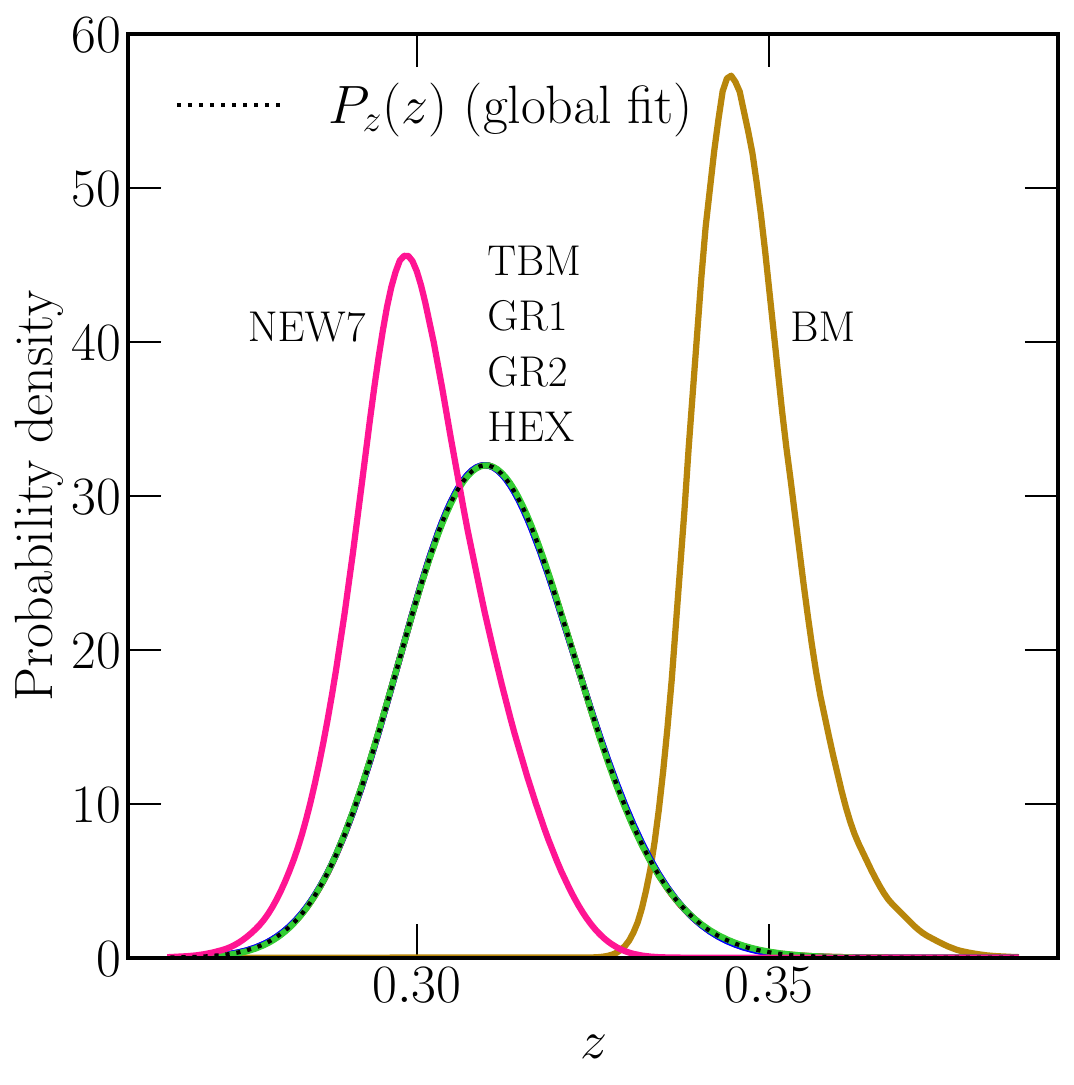}
	\end{minipage}
	\caption{
		Left: colored solid lines for $P_c(c)$ as estimated based on $P_{GC}$ , while dashed black lines are for the
		distribution obtained by marginalizing $c$ from $P_{b,c}(b,c)$.
		Right: $P_z(z)$ as obtained from
		$P_c(c)$. The dotted black line is $P_z(z)$
		based on the global fit.
		\label{fig:PcPzbm}
	}
\end{figure}

We now have $P_z(z)$ given by
\begin{align}
P_z(z) & = \int db\, dc\, \deltafun\left(\eval[2]{f}_{a=\frac{x_0}{b}} - z\right)
P_{b,c}(b,c) \nonumber\\
& = \int db\, \frac{b(1 - x_0)}{2\sqrt{(1 - b)(b - x_0)(1 - z_0)x_0
z_0}}\,  P_{b,c}(b,c_0(b, z)),
\end{align}
where $c_0(b,z)$ is given by Eq.~\eqref{eq:thec0}.
Similarly, we can write 
\begin{align}
P_{\cos\delta}(\cos\delta) & = \int da\, db\, dc\, \delta(\tilde{g} - \cos\delta)
	P_{a|b}(a) P_{b,c}(b,c)\\
& = \int dc\,
	\eval{\left[\left(\frac{\partial \tilde{g}}{\partial b}\right)^{-1}
	P_{b,c}(b,c)\right]}_{b\text{ such that } \tilde{g} = \cos\delta},
\label{eq:PcosintfinalPbc}
\end{align}
with $P_{a|b}(a)$ given in Eq.~\eqref{eq:PabPba}.
The resulting probability distribution $P_{\cos\delta}(\cos\delta)$ for all the patterns
is shown in Figure~\ref{fig:Pcosdelta} (black dashed lines). For illustrative purposes, we compare these results with those that are obtained
in the simplified case that the model parameter distributions are taken to be independent (solid colored lines).
While for the TBM and GR2 patterns the distributions do not change noticeably,
the HEX patterns show a larger spread when integrating with $P_{b,c}(b,c)$.
This difference is related to the inaccuracy in the estimation of $P_c(c)$
as shown in the left panel of Figure~\ref{fig:PcPzbm}, in which we compare the results for the estimation of 
$P_c(c)$ using $P_{GC}$ (solid colored lines)
and marginalizing $b$ out of $P_{b,c}(b,c)$ (dashed black lines).
It is expected that the inaccuracy of assuming that $P_{b|c}(b) =
P_b(b)$, should affect the resulting estimation of $P_c(c)$.

Clearly, more care must be taken for the cases of BM and NEW7 mixing.
As is well known, the viable parameter space for these two cases
is significantly smaller than it is for the other mixing patterns considered here,
as BM mixing generically predicts values for the solar mixing angle that are quite large
while for NEW7 the prediction is on the smaller side,
falling on the tail ends of the experimentally allowed region.
For instance, inside the domain of allowed values of $b$ and $c$,
the allowed combinations of $y$ and $z$ are far more limited
in the BM and NEW7 patterns than for the other patterns,
disallowing large swathes of parameter space.
This lack of parameter space freedom
results in the fact that the step of approximating $P_{b|c}(b)$ by $P_b(b)$,
as given by Eq.~\eqref{eq:PbPy},
is far more inaccurate for BM and NEW7 mixing
than for the other four scenarios.
Similarly, we see that $P_b(b)$ for BM mixing
does not correspond to Eq.~\eqref{eq:PbPy}
as well as it does for the other patterns,
as can be seen on the right side of Figure~\ref{fig:zlines}.
Lastly, we see that the scenarios BM and NEW7 mixing
are completely unable to reproduce the global fit of $P_z(z)$, (Figure~\ref{fig:PcPzbm}, right side),
which indicates the unsuitability of determining $P_c(c)$ by optimizing $P_z(z)$ to match the global fit.

\begin{figure}[h]
\includegraphics[scale=\imgscale]{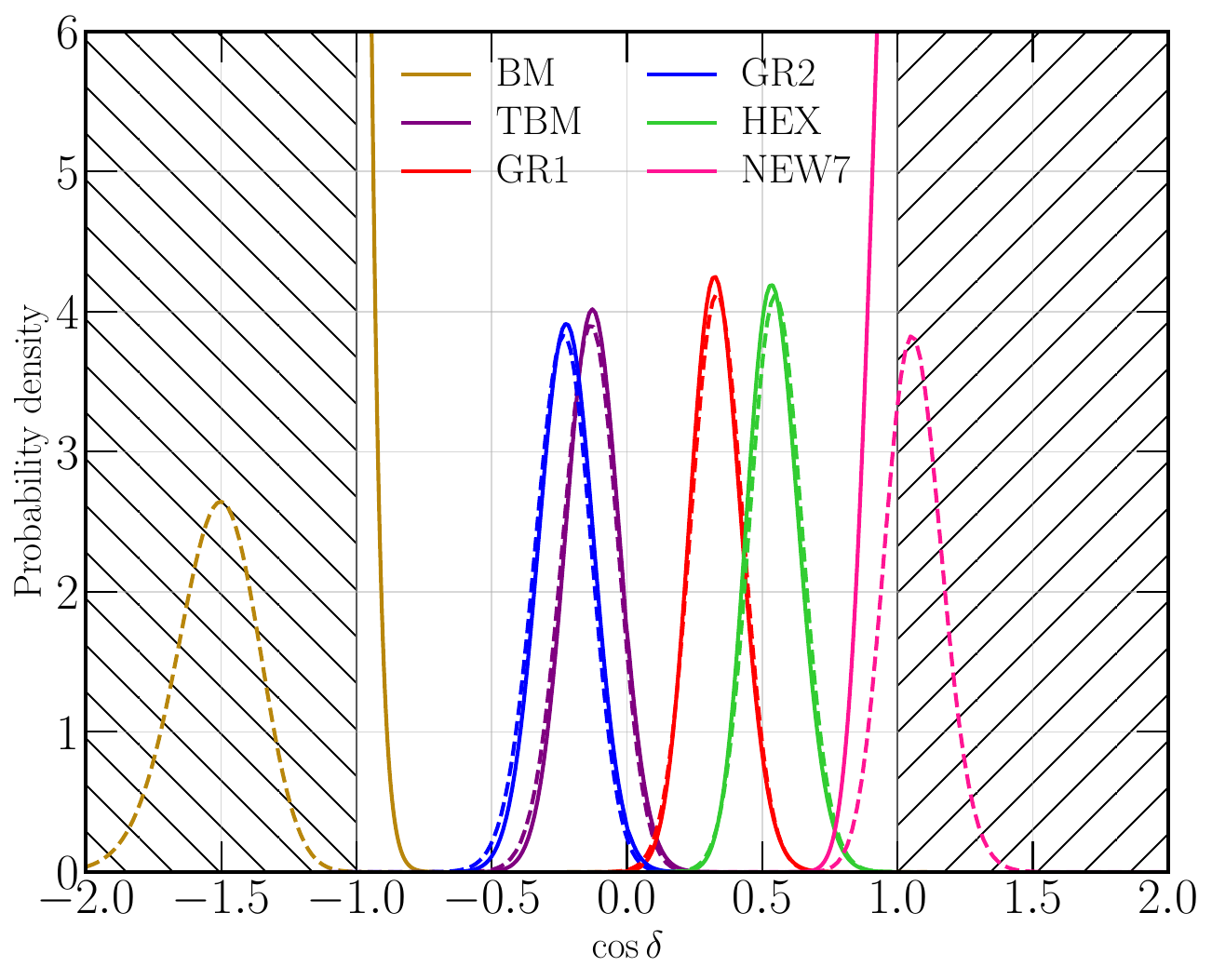}
\caption{The distribution for $P_{\cos\delta}(\cos\delta)$ as given in Figure~\ref{fig:Pcosdelta}, overlaid with $P_{\cos\delta}(\cos\delta)$ as obtained from treating the experimental distributions as uncorrelated inputs (depicted with dashed lines). 
}
\label{fig:compareking2}
\end{figure}


Furthermore, we can also see this feature when comparing our approach for determining $P_{\cos\delta}(\cos\delta)$ as compared to approaches taken in previous literature, in which the experimental distributions are taken to be uncorrelated inputs in determining the posterior probability distribution. An assumption that is often made in the literature is that the model can always accommodate all values of the experimentally measured parameters. 
However, we have seen that in this particular (quite simple) model scenario, the range of $z$ is limited. 

In Figure~\ref{fig:compareking2}, we show a comparison of $P_{\cos\delta}(\cos\delta)$ as obtained using our methods to the $P_{\cos\delta}(\cos\delta)$ as calculated using Eq.~(\ref{eq:sumruleorig}), where each experimental distribution is assumed to be uncorrelated with the others. Clearly, the BM and NEW7 cases show the significant difference in these methods.
The shift of the peak to unphysical regions for the case in which the experimental observables are taken to be uncorrelated
reflects the well-known fact that the BM pattern has difficulty reproducing the best fit values of the experimental data,
and the NEW7 pattern has similar difficulties.
Here we also note that the remaining patterns show a very slight shift in the peaks in the distributions for $P_{\cos\delta}(\cos\delta)$ as well.
This is as expected since the theoretically allowed values for the observable parameters are also restricted,
albeit not as drastically as in the BM pattern.
We also see a slight shift in our resulting distributions when compared to previous approaches,
in part due to the use of $\chi^2$ obtained via global fits.
We further note that with the method used here, the predicted distribution for $\cos\delta$ automatically only falls in the physical range, as expected. 


\section{Conclusions}
\label{sec:conclusions}
In this paper, we investigated the predictions of the sum rule of Eq.~(\ref{eq:sumruleorig}) for the leptonic Dirac CP-violating phase $\delta$, using a procedure that ensures unitarity of the lepton mixing matrix at all stages. We have taken as an example a specific, well-studied set of theoretical models with a single $1-2$ charged lepton rotation with a single source of CP-violation, that acts on the TBM, BM, GR1, GR2, HEX, or NEW7 (unperturbed solar mixing angle of $\pi/7)$  form of the neutrino mixing matrix at leading order.  Here we have made the nontrivial assumption that such simple models can provide a correct description of lepton mixing, with no non-standard interactions or additional exotic fields, as might be expected in a more complete theoretical description of lepton mass matrices.  

We have used mixing angle data to inform our understanding of the probability distributions of the continuous model parameters in this simple class of models, and further assumed that the precisely measured reactor mixing angle distribution can be approximated as a delta-function distribution.   
We calculated the predicted $\cos\delta$ distribution in these scenarios, with a procedure that guarantees that the values of $\cos\delta$ will always lie solely within the physically allowed region.  The peaks in these distributions show small but nonzero shifts from previous results in the literature for the scenarios in which the model parameters can generally access wide regions of the experimentally allowed ranges of the lepton mixing angles (the TBM, GR1, GR2, and HEX cases), but there is a significant shift for cases like BM, and to a lesser extent NEW7, where the theoretical constraints yield only limited access to the full experimentally allowed atmospheric and solar mixing angle ranges.  

Here we have illustrated this approach for a long-known and very well-studied class of very simple toy models, and found that our procedure only leads to small differences in the $\cos\delta$ distribution compared to prior approaches, except in cases where the model itself only allows a very limited parameter range that can reproduce the lepton mixing angle data.  However, we emphasize that refinements in the procedure for predicting distributions for $\cos\delta$ to ensure that the theoretical constraints in a given set of theoretical models are satisfied, such as those described here, will necessarily become more timely and important in the future as the measurements of the lepton mixing angles continue to improve. Thus, we believe that these issues are worthy of consideration in future analyses, as we continue to work toward the long-standing goal of resolving aspects of the Standard Model flavor puzzle.


\section*{Acknowledgments}


We thank I.~M.~Lewis for helpful comments and suggestions.  The work of L.L.E. and A.B.R is supported by the U.S.~Department of Energy under contract number DE-SC0017647. A.J.S. would like to acknowledge partial support from CONACYT project CB2017-2018/A1-S-39470 (M\'exico).  R.R. is grateful for the hospitality of FdeC-CUICBAS  Universidad de Colima where part of this work was carried out. The
work of R.R. is supported in part by the Ministry of Science and Technology (MoST)
of Taiwan under grant numbers 107-2811-M-001-027 and 108-2811-M-001-550.

\appendix

\section{Appendix}
In what follows, we summarize several useful facts about the distribution functions used in this work.
We begin by referring to the usual normalized Gaussian density function as
\begin{align}
\label{eq:gaussnorm}
P_{\rm Gauss}(\alpha; \mu, \sigma) = \frac{1}{\sqrt{2\pi}\sigma}\exp\!\left[
	-\frac{1}{2}\left(\frac{\alpha - \mu}{\sigma}\right)^2,
\right]
\end{align}
where $\mu$ represents the mean and $\sigma$ the standard deviation.

This is an obvious choice of distribution function to use in our analysis.  However, nothing ties us to this choice and one could consider distribution
functions that deviate from Eq.~\eqref{eq:gaussnorm} through their dependence on additional
parameters.
Two well-known examples of such distribution functions are given by
\begin{align}
\label{eq:gaussskew}
P_{\rm skew}(\alpha; \mu, \sigma, s) & = P_{\rm Gauss}(\alpha; \mu, \sigma)\left[
	1 + \erf\left(\frac{s}{\sqrt{2}}\cdot\frac{\alpha - \mu}{\sigma}\right)
\right],\\
\label{eq:gaussgc}
P_{\rm GC}(\alpha; \mu, \sigma, s, k) & = P_{\rm Gauss}(\alpha; \mu, \sigma)\left[
	1 + \frac{s}{6\sqrt{2}}H_3\left(\frac{\alpha - \mu}{\sqrt{2}\sigma}\right)
	+ \frac{k}{96}H_4\left(\frac{\alpha - \mu}{\sqrt{2}\sigma}\right)
\right],
\end{align}
where ${\erf(\beta)}$ is the error function and $H_{3,4}(\beta)$ are
(physicists') Hermite polynomials, {\it i.e.}, $H_3(\beta) = 8\beta^3 - 12\beta$ and
$H_4(\beta) = 16\beta^4 - 48\beta^2 + 12$.
A choice of values for the parameters after the semicolon ($\mu$, $\sigma$, $s$, $k$, etc.) defines a
probability distribution for the placeholder variable $\alpha$.
The distribution defined in Eq.~\eqref{eq:gaussskew} is known as the
\emph{skew normal distribution} since it adds a parameter, $s$, that controls
the skewness of the distribution.
Eq.~\eqref{eq:gaussgc} is called the Gram-Charlier distribution which,
additional to the skewness parameter, adds a kurtosis parameter, $k$, to
parameterize deviations on the tails of the distribution.


 
\end{document}